\documentclass[]{x2lab}

\usepackage{microtype}
\usepackage{graphicx}
\usepackage{subcaption}
\usepackage{tikz}
\usetikzlibrary{arrows.meta,positioning,calc,decorations.pathreplacing}
\usepackage{csquotes}
\usepackage{afterpage}
\usepackage{placeins}

\usepackage{amsmath}
\usepackage{amssymb}
\usepackage{mathtools}
\usepackage{amsthm}
\usepackage{xspace}
\usepackage{colortbl}
\usepackage{multirow}
\usepackage{enumitem}
\usepackage{wrapfig}
\usepackage{nicefrac}
\usepackage{makecell}
\usepackage{multirow}
\usepackage{xcolor}
\usepackage{booktabs}
\usepackage{algorithm}
\usepackage{algpseudocode}
\usepackage{listings}
\usepackage{xcolor}
\usepackage{pifont}

\definecolor{fbApp}{HTML}{c8e7fa}
\definecolor{fbPurple3}{HTML}{f0ebf5}

\definecolor{citecolor}{HTML}{0071BC}
\definecolor{linkcolor}{HTML}{ED1C24}

\definecolor{citecolor}{HTML}{0071BC}
\definecolor{linkcolor}{HTML}{ED1C24}

\title{DMuon: Efficient Distributed Muon Training with Near-Adam Overhead}
\author[]{X SQUARE ROBOT TEAM}

\abstract{
Matrix-orthogonalization-based optimizers, exemplified by Muon, have demonstrated strong convergence behavior across a wide range of modern deep learning workloads. The matrix-aware updates offer a compelling alternative to conventional element-wise optimization, particularly as model architectures continue to grow in scale and heterogeneity. Yet contemporary distributed training infrastructure built around the assumption of element-wise optimizers is poorly matched to matrix-level optimizers such as Muon, whose updates couple entire weight matrices and require costly Newton-Schulz iterations. Vanilla Muon implementations incur \textbf{more than} $\mathbf{2\times}$ the cost of forward and backward passes. To close this gap, we present DMuon, an open-source distributed Muon implementation that integrates into existing training pipelines as a drop-in module, with no framework-level modifications. Across both embodied foundation model and large language model (LLM) training workloads, DMuon achieves a $\mathbf{1.48\times\!\text{--}\!3.01\times}$ speedup in end-to-end step time and a $\mathbf{6.85\times\!\text{--}\!163.00\times}$ speedup in optimizer-step time, bringing per-step latency to \textbf{near-AdamW} levels and enabling efficient scaling in our model training.
}
  
\date{\today}

\metadata[Code]{\url{https://github.com/X-Square-Robot/dmuon}}

\begin{document}
\maketitle

\section{Introduction}
\label{sec:intro}

\begin{figure}[t]
\centering

\begin{subfigure}[t]{0.8\textwidth}
\begin{lstlisting}[language=Python]
import (*\textbf{dmuon}*)
(*\textbf{dmuon}*).dedicate_params(model, mesh)
opt = (*\textbf{dmuon}*).Muon(model)
\end{lstlisting}
\caption{Programming interface}
\label{subfig:drop-in}
\end{subfigure}

\vspace{0.3cm}

\begin{subfigure}[t]{0.48\textwidth}
    \centering
    \includegraphics[width=\linewidth]{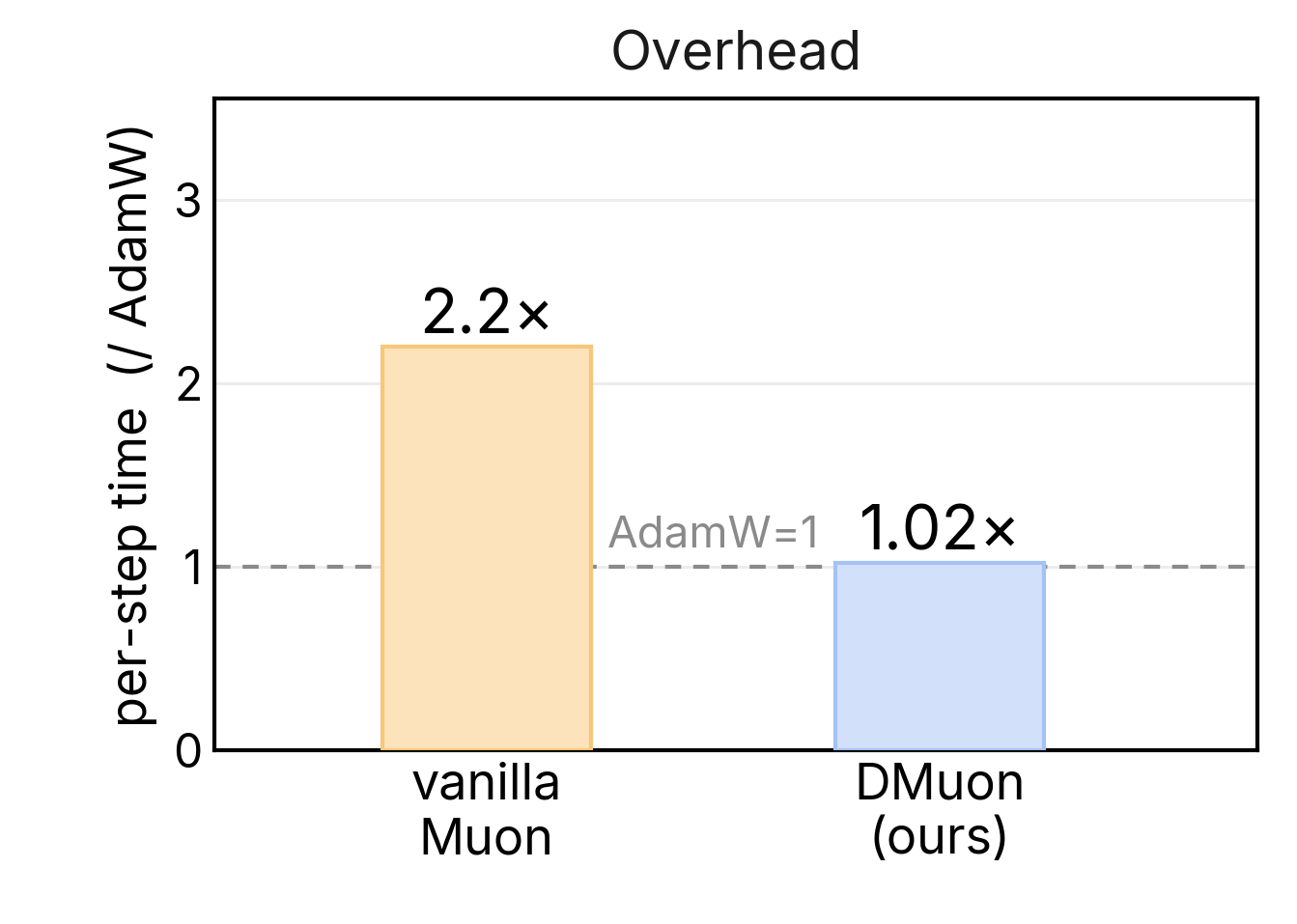}
    \caption{Negligible per-step overhead}
    \label{subfig:overhead}
\end{subfigure}
\hfill
\begin{subfigure}[t]{0.48\textwidth}
    \centering
    \includegraphics[width=\linewidth]{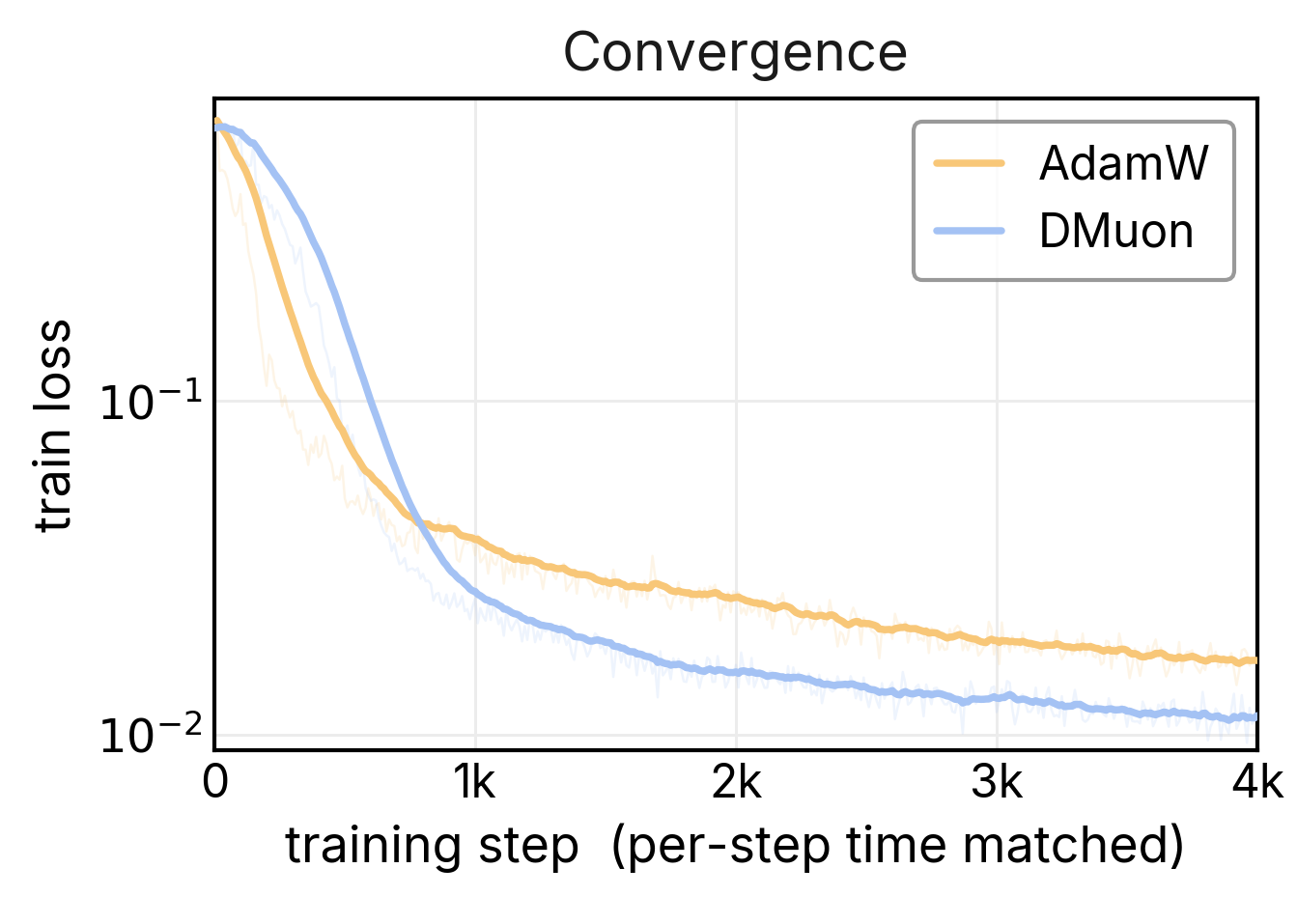}
    \caption{Faster wall-clock convergence}
    \label{subfig:convergence}
\end{subfigure}

\vspace{0.2cm}
\caption{(a) Three lines of code replace AdamW with
 DMuon, at a (b) negligible per-step overhead, yielding
(c) faster wall-clock convergence. The 3.1$\times$ speedup is averaged across the four models in \Cref{sec:eval-main}, and the loss curves are captured from our real training run of \textsc{Wall-OSS-0.5}.}
\label{fig:teaser-fig}

\end{figure}

Large-model training has begun to move beyond the element-wise optimization
paradigm that has dominated since
AdamW~\cite{loshchilov2017adamw}. Matrix-aware optimizers, most prominently
Muon~\cite{jordan2024muon}, apply a Newton--Schulz iteration to the
momentum-aggregated gradient of each weight matrix, producing updates whose
singular values are driven toward a near-uniform spectrum.
Recent works suggest that this update geometry
translates into meaningful practical benefits, Moonlight
~\cite{moonlight2025}
reports approximately $2\times$ higher compute efficiency than AdamW under
compute-optimal pretraining, and Kimi-K2, DeepSeek-V4
~\cite{kimi-k2,deepseekai2026deepseekv4} have adopted Muon in production-scale
training, providing further evidence that matrix-aware optimization is viable
at frontier-model scale. The algorithmic promise of matrix-aware optimizers
is therefore increasingly well established. Their deployment cost, however, remains a significant practical obstacle.

Modern distributed training stacks execute the optimizer step on
\emph{shards} of each parameter, one shard per rank. The implicit contract is
that the optimizer's update rule is element-wise: AdamW's updates apply to a
shard with no knowledge of any other shard. Muon breaks this contract. Its
core computation, the Newton--Schulz iteration, operates on the full weight
matrix rather than individual shards. Before a Muon update can be computed,
the matrix must therefore be reconstructed, introducing communication that is
absent from conventional element-wise optimizers.
This additional communication is not a minor overhead. It appears at every
optimizer step, for every matrix parameter, and scales with model size and
distributed width. In practice, the cost of distributed Muon can rival or
even exceed the combined wall-clock time of the forward and backward passes.

The extra per-step cost erodes Muon's FLOP-level advantage and weakens the
wall-clock incentive to deploy it at scale. The penalty is particularly
pronounced for embodied-model training. Compared to contemporary LLM and VLM
pretraining~\cite{gpt4,qwen25,qwen25vl,paligemma}, whose forward--backward passes
typically involve substantially more computation per optimizer step, VLA
training~\cite{pi0,pi05,openvla,wallx,walloss05,lingbotvla} typically uses
much shorter temporal contexts. As a result, the forward--backward phase
occupies a much smaller fraction of the step time, making optimizer overhead
significantly harder to amortize.

Therefore, we present DMuon, a distributed Muon that closes
the per-step performance gap to AdamW. Concretely, this work makes
the following contributions:

\begin{itemize}

\item \textbf{Fine-grained communication optimization.}
We develop an owner-centric strategy that minimizes the communication overhead
of distributed Muon execution while preserving exact optimizer semantics.

\item \textbf{Shape-adaptive execution stack.}
We develop a high-performance execution engine that combines batched Gram Newton–Schulz execution, symmetry-aware kernels, and DSL-driven autotuning, enabling
efficient orthogonalization across heterogeneous matrix workloads.

\item \textbf{Computation-aware load balancing.}
We formulate owner assignment as a measured-cost optimization problem,
leveraging runtime profiling to balance work across owners under batching,
kernel-selection, and autotuning effects.

\item \textbf{Drop-in distributed module.}
We provide DMuon as a drop-in module that remains orthogonal to the host
training framework and parallelism strategy. Existing training pipelines can
benefit from distributed Muon with minimal integration effort while
preserving exact optimizer semantics.

\item \textbf{Production-scale validation.}
We validate DMuon on both robotics and language-model training workloads,
including the production training of WALL-OSS-0.5 and WALL-WM, demonstrating
near-AdamW optimizer overhead while retaining Muon's convergence benefits.
\end{itemize}

In sum, as shown in Figure~\ref{fig:teaser-fig}, DMuon enables Muon's convergence benefits to translate into practical
training speedups. The following sections describe how DMuon combines communication, scheduling, and kernel optimizations to achieve these gains.

\section{Background and Motivation}
\label{sec:background}

This section establishes the three pieces of context needed to motivate DMuon's
design: the Muon optimizer and its Newton-Schulz iteration
(\S\ref{sec:bg-muon}), the sharded-training abstractions DMuon operates under
(\S\ref{sec:bg-sharding}), and the structural mismatch between the two
(\S\ref{sec:bg-mismatch}) that drives the system cost we set out to eliminate.

\subsection{Muon and the Newton--Schulz Iteration}
\label{sec:bg-muon}

Muon~\cite{jordan2024muon} is a matrix-aware optimizer designed for
two-dimensional parameters such as the weight matrices of linear and
attention projections. For a weight matrix
$W \in \mathbb{R}^{m\times n}$ with momentum-smoothed gradient $M_t$, the
core update rule is

\begin{equation}
W_{t+1}
=
W_t
-
\eta \cdot \mathrm{NS}_k(M_t),
\label{eq:muon-update}
\end{equation}

where $\mathrm{NS}_k(\cdot)$ denotes $k$ steps of a Newton--Schulz iteration
that approximates the matrix sign function, equivalently the orthogonal
factor $UV^\top$ of the SVD
$M_t = U \Sigma V^\top$. Each Newton--Schulz step has the form

\begin{equation}
X_{i+1}
=
aX_i
+
b\,X_iX_i^\top X_i
+
c\,(X_iX_i^\top)^2X_i,
\label{eq:ns-step}
\end{equation}

where the coefficients $(a,b,c)$ are chosen to drive the singular values of
$X_i$ toward unity. In practice, $k=5$ iterations are typically sufficient,
and the computation is performed in reduced precision since only the singular
subspace, rather than the singular values themselves, must be preserved.

The practical cost of Muon is dominated by the Newton--Schulz iteration. A
single optimizer step requires multiple Newton--Schulz updates, each
consisting primarily of large matrix multiplications. Consequently, the cost
of orthogonalization grows rapidly with matrix dimensions and can become a
substantial component of the overall training step.

As model sizes continue to increase, efficiently executing the
Newton--Schulz iteration becomes critical for practical deployment. The
remainder of this paper focuses on reducing the systems overhead associated
with this computation.

\subsection{Sharded Training Abstractions}
\label{sec:bg-sharding}

Distributed training systems partition weight matrices in different ways.
The location of shard boundaries determines how a matrix is stored,
communicated, and reconstructed during training. We briefly review the three
sharding schemes most relevant to matrix-aware optimization: ZeRO-style
sharding, FSDP-style sharding, and tensor parallelism.

\paragraph{ZeRO-style: flatten then shard.}
ZeRO~\cite{rajbhandari2020zero} (as realized by
DeepSpeed~\cite{rasley2020deepspeed}) first
flattens a group of parameters into a single contiguous buffer and then
partitions that buffer across data-parallel ranks. Shard boundaries are
defined by storage layout rather than tensor structure, so a single matrix
may span multiple ranks and share a shard with unrelated parameters.

\paragraph{FSDP: per-tensor sharding.}
FSDP2~\cite{zhao2023fsdp} represents each parameter as a DTensor sharded
along its leading dimension. Unlike ZeRO-style sharding, partition boundaries
align with tensor structure, making reconstruction and communication local to
an individual parameter. HSDP extends the same abstraction to a two-dimensional
$D_{\mathrm{shard}}\times D_{\mathrm{replica}}$ mesh to reduce communication
distance in multi-node deployments.

\paragraph{Tensor parallelism: shard along a semantic axis.}
Tensor parallelism (TP)~\cite{shoeybi2019megatron} partitions weight matrices
along model-semantic dimensions such as attention heads or MLP hidden
channels. These partitions persist throughout execution and therefore become
part of the model definition itself rather than merely a storage strategy.

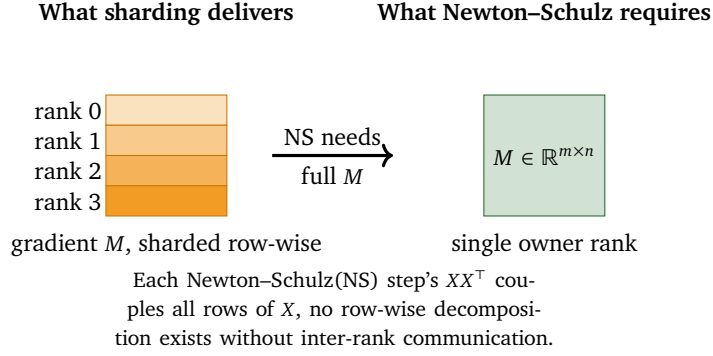
\begin{figure}[h]
\centering
\definecolor{dmuongreen}{HTML}{2E7D32}   
\definecolor{baseorange}{HTML}{F08A00}   
\begin{tikzpicture}[font=\small, node distance=4mm]
  \node[anchor=south] at (0, 2.2) {\textbf{What sharding delivers}};
  \foreach \i/\op in {0/25, 1/45, 2/65, 3/85} {
    \fill[baseorange!\op] (-0.8, 1.4 - 0.4*\i) rectangle (0.8, 1.0 - 0.4*\i);
    \draw[baseorange!85!black, line width=0.4pt]
      (-0.8, 1.4 - 0.4*\i) rectangle (0.8, 1.0 - 0.4*\i);
    \node at (-1.3, 1.2 - 0.4*\i) {rank \i};
  }
  \node[anchor=north] at (0, -0.3) {gradient $M$, sharded row-wise};

  \draw[->, very thick] (1.4, 0.6) -- (3.0, 0.6)
    node[midway, above, align=center] {NS  needs}
    node[midway, below, align=center] {full $M$};

  \node[anchor=south] at (5, 2.2) {\textbf{What Newton–Schulz requires}};
  \fill[dmuongreen!22] (4.2, 1.4) rectangle (5.8, -0.2);
  \draw[dmuongreen!85!black, line width=0.4pt] (4.2, 1.4) rectangle (5.8, -0.2);
  \node at (5, 0.6) {$M \in \mathbb{R}^{m\times n}$};
  \node[anchor=north] at (5, -0.3) {single owner rank};

  \node[anchor=north, align=center, text width=8cm] at (2.2, -0.8)
    {\footnotesize Each Newton–Schulz(NS) step's $X X^\top$ couples all rows of $X$, no
      row-wise decomposition exists without inter-rank communication.};
\end{tikzpicture}
\caption{The granularity mismatch. Sharded training partitions each weight
matrix and its gradient across ranks, but the Newton-Schulz iteration
(Eq.~\ref{eq:ns-step}) requires the full matrix on a single device to compute
$X X^\top$.}

\label{fig:granularity-mismatch}
\end{figure}

\subsection{Problems and Opportunities}
\label{sec:bg-mismatch}
\textbf{Granularity mismatch.} Given the sharding schemes in \S\ref{sec:bg-sharding}, the conflict with
Muon is immediate: training systems expose matrix parameters in pieces, while
Muon must run Newton--Schulz on a full reduced matrix gradient. As illustrated in Figure~\ref{fig:granularity-mismatch}, for a matrix
parameter $W\in\mathbb{R}^{m\times n}$, Muon requires
$M=\frac{1}{D}\sum_{r=1}^{D} g_r$ before applying
Eq.~\ref{eq:ns-step}. Currently, a distributed system must therefore materialize the full matrix somewhere before
running the Newton--Schulz iteration.

A straightforward way to satisfy this full-matrix requirement is
\emph{gather-then-compute}. Each rank materializes the full
reduced matrix gradient, runs the Newton--Schulz iteration locally, and then
keeps the portion of the resulting update corresponding to its local parameter
shard. This strategy is simple and preserves Muon's update rule, but it
exposes the two costs that an efficient distributed system must address:
matrix materialization and redundant orthogonalization.

\textbf{Matrix materialization}. At every
optimizer step, the full reduced matrix must be reconstructed from its
distributed representation. This introduces optimizer-specific collective
communication whose volume scales with matrix size and distributed width. The
cost is especially pronounced because Muon operates at the level of
individual matrices: each matrix must be made available as a coherent object
before the Newton--Schulz iteration can begin.

\textbf{Replicated orthogonalization}. Once every rank has
materialized the same matrix, the Newton--Schulz iteration is executed
identically on all ranks. Aggregated optimizer compute is therefore multiplied
by the number of ranks that redundantly run the same orthogonalization, even
though each matrix only needs to be orthogonalized once. Since this iteration is already Muon's primary computational bottleneck, such redundant execution can cause the optimizer step to rival or exceed the combined cost of the forward and backward passes at scale.

Intuitively, the replication problem can be alleviated through
owner-like execution, where each matrix is assigned to a
designated owner.  Yet removing
redundant computation alone is not enough: communication and execution
overheads quickly become the new bottlenecks. DMuon addresses these through a
multi-level co-designed architecture, forming an
integrated solution for efficient distributed Muon training  (\S\ref{sec:design}).      
\section{System Design}
\label{sec:design}

Our goal is to make Muon practical in sharded distributed training while preserving its exact update rule, and simultaneously reduce its per-step overhead to a near-AdamW
level. We address this challenge by employing system-level design rather than isolated
optimizations: DMuon orchestrates runtime, scheduling, and kernel-level techniques
to mitigate the granularity mismatch identified in \S\ref{sec:bg-mismatch}.  In this section, we present the overall system design of DMuon.

\begin{figure}[t]
\centering
\includegraphics[width=\linewidth]{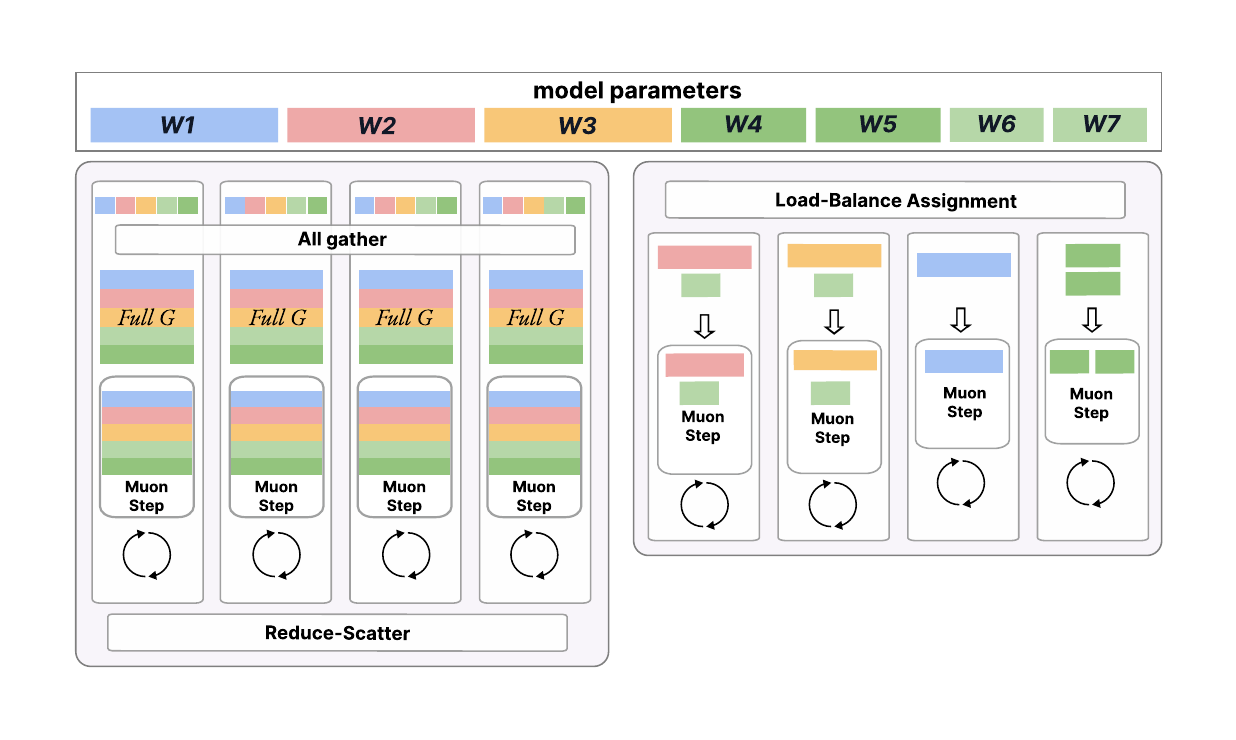}
\caption{DMuon assigns one owner rank per matrix parameter. \emph{Bottom-left, FSDP sharding:} every rank
all-gathers the full gradient $G$ of every matrix and runs the same Muon
step locally, redundant compute scales with the data-parallel width.
\emph{Bottom-right, Load balance assignment:} each matrix is assigned to a single owner, only the owner runs the
Muon step, other ranks idle on that
matrix. }
\label{fig:design-overview}
\end{figure}

\subsection{Design Overview}
\label{sec:design-overview}

Figure~\ref{fig:design-overview} contrasts the gather-then-compute baseline
with DMuon's owner-style execution. In the baseline, each rank materializes
the full gradient of every matrix parameter and runs the same Newton--Schulz
iteration locally, so optimizer compute is replicated across the data-parallel
group. DMuon instead assigns each matrix parameter to a single \emph{owner}
rank. The owner maintains the authoritative optimizer state for that matrix
and computes its Muon update once, while non-owner ranks materialize the parameter
only when it is needed by forward or backward computation.

This change removes redundancy, but it also turns distributed
Muon into an owner-side execution problem: gradients must be routed to owners,
updated parameters must be made visible to non-owners, owner workloads must be
balanced, and the remaining Newton--Schulz computation must run efficiently.
DMuon addresses these requirements through a coordinated system design rather
than a single kernel or communication primitive.

At a high level, one DMuon training step proceeds in four phases:

\begin{enumerate}
\item \textbf{Parameter materialization.}
Before a layer executes, the matrix parameters required by that layer are
materialized from their owners into packed buffers consumed by the forward
computation. This makes owner-held parameters appear to the model as ordinary
local tensors during execution.

\item \textbf{Gradient routing.}
During backward, gradients are accumulated into the same packed buffers and
then reduced to the corresponding owners. The reduction produces the same
averaged full-matrix gradient that a synchronous Muon reference would use,
but materializes it only at the owner.

\item \textbf{Owner-side Muon update.}
Each owner runs the Muon update for its assigned matrices and applies the
result to its local authoritative copy. And non-matrix parameters continue to follow
the optimizer path provided by the host training stack.

\item \textbf{Asynchronous publication.}
After the owner updates its parameters, the new values are published back to
the ranks that will need them in the next step. This publication is scheduled
asynchronously so that its cost can be overlapped with useful work whenever
possible.
\end{enumerate}

The remainder of this section expands these pieces. We first describe the
owner runtime and communication flow (\S\ref{sec:design-schedule}), then the
efficient owner-side Muon execution path (\S\ref{sec:design-gram-ns}), the
measurement-based owner assignment (\S\ref{sec:loadbalance}), and finally
the full step-level pipeline (\S\ref{sec:design-pipeline}).

\subsection{Owner Centric Communication Optimization}
\label{sec:design-schedule}

Under the owner-style execution, each matrix update is computed once by an
owner rank, and the updated parameter must be made available to the ranks
that execute subsequent forward and backward computation. This removes
replicated Newton--Schulz computation, but also gives owner communication a
distinct asymmetric structure. In the forward pass, updated parameters are
published from one owner to many consumers. In the backward pass, gradients
from many ranks are reduced back to the owner. This asymmetry creates an
optimization opportunity beyond simply replacing one collective with another.

We exploit this opportunity with a communication runtime tailored to the
owner-style access pattern. Rather than treating parameter publication and
gradient routing as uniform all-rank collectives, the runtime organizes them
around the hierarchy of the training mesh and schedules them to expose overlap
with computation. The remainder of this section shows how this communication
structure is mapped onto the training mesh and pipelined across the forward
and backward passes to reduce effective communication overhead.

\begin{figure}[t]
\centering
\includegraphics[width=\linewidth]{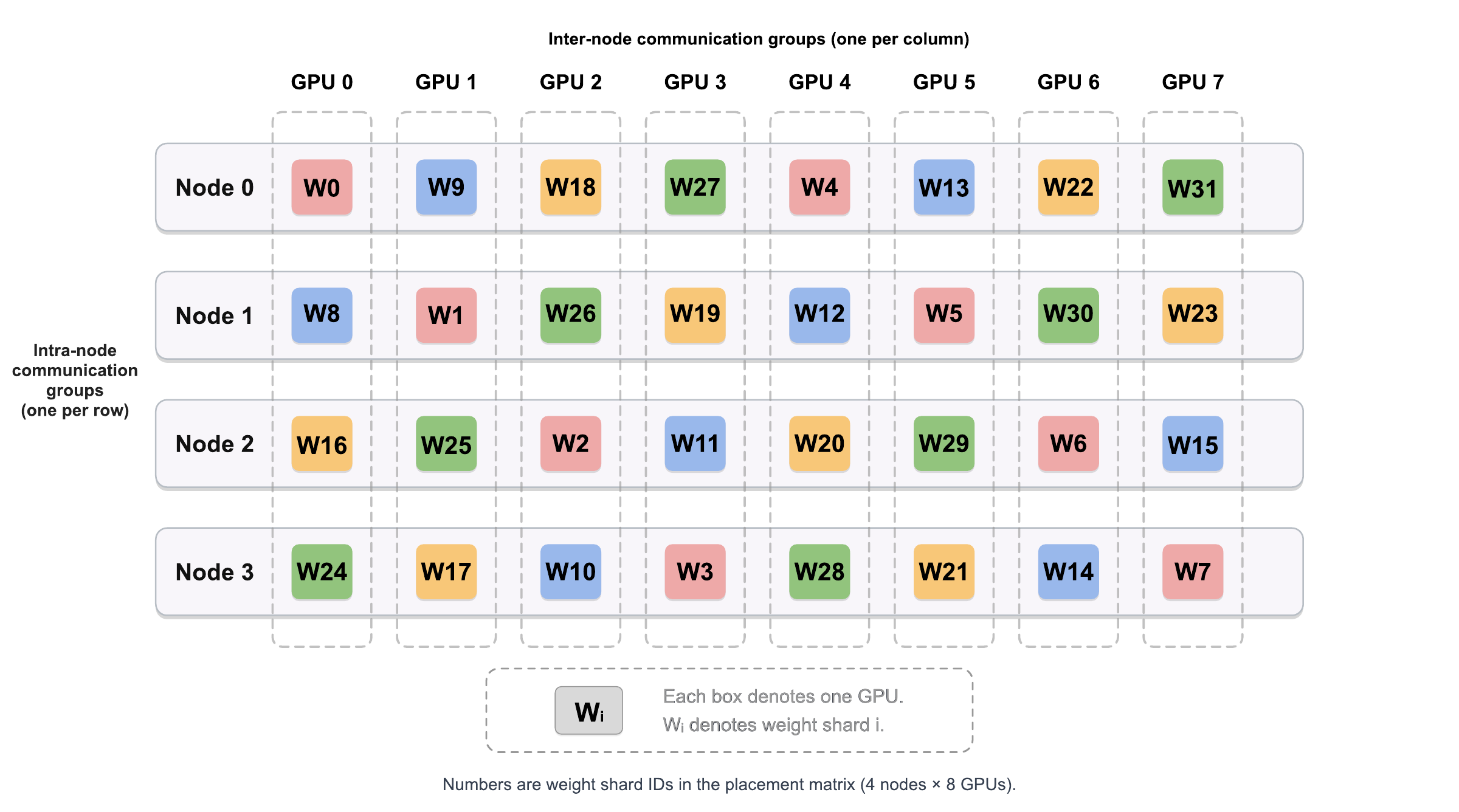}
\caption{\textbf{Balanced shard layout} on a $4\times8$ example mesh (node $\times$
GPU). A closed-form XOR rule assigns each GPU one weight shard, columns
are inter-node communication groups, rows intra-node groups (colors = blocks
of eight consecutive IDs). The rule balances ownership across both dimensions
while preserving locality.}
\label{fig:weight_layout}
\end{figure}

\subsubsection{Fine-Grained Weight Layout}

We reorganize both owner-to-all broadcasts and all-to-owner reductions into a two-stage hierarchy comprising intra-node and inter-node communication. The communication schedule is effective only when concurrent collectives avoid repeatedly contending for the same communication group. We therefore use a fine-grained
owner-slot layout that disperses nearby matrix communication units across GPU
columns while rotating their owner nodes across consecutive groups.

For a $4\times 8$ deployment with four nodes and eight GPUs per node, let
$w$ denote the logical index of a matrix in the communication schedule. We map
it to an owner slot by

\begin{equation}
\mathrm{gpu}(w) = w \bmod 8,
\qquad
\mathrm{node}(w) =
(w \bmod 4) \oplus
\left(\left\lfloor \frac{w}{8} \right\rfloor \bmod 4\right),
\label{eq:comm-layout}
\end{equation}

where $\oplus$ denotes bitwise XOR. The GPU assignment distributes consecutive
matrices across the eight inter-node columns, while the XOR term rotates the
owner node across groups of eight matrices. As a result, a lookahead window of
matrix publications is spread over different inter-node communication groups
instead of being concentrated on a single column. This layout reduces
collective-level contention and enables the forward and backward pipelines
described next. Figure~\ref{fig:weight_layout} illustrates the mapping on a
$4\times 8$ mesh.

\begin{figure}[t]
\centering
\includegraphics[width=\linewidth]{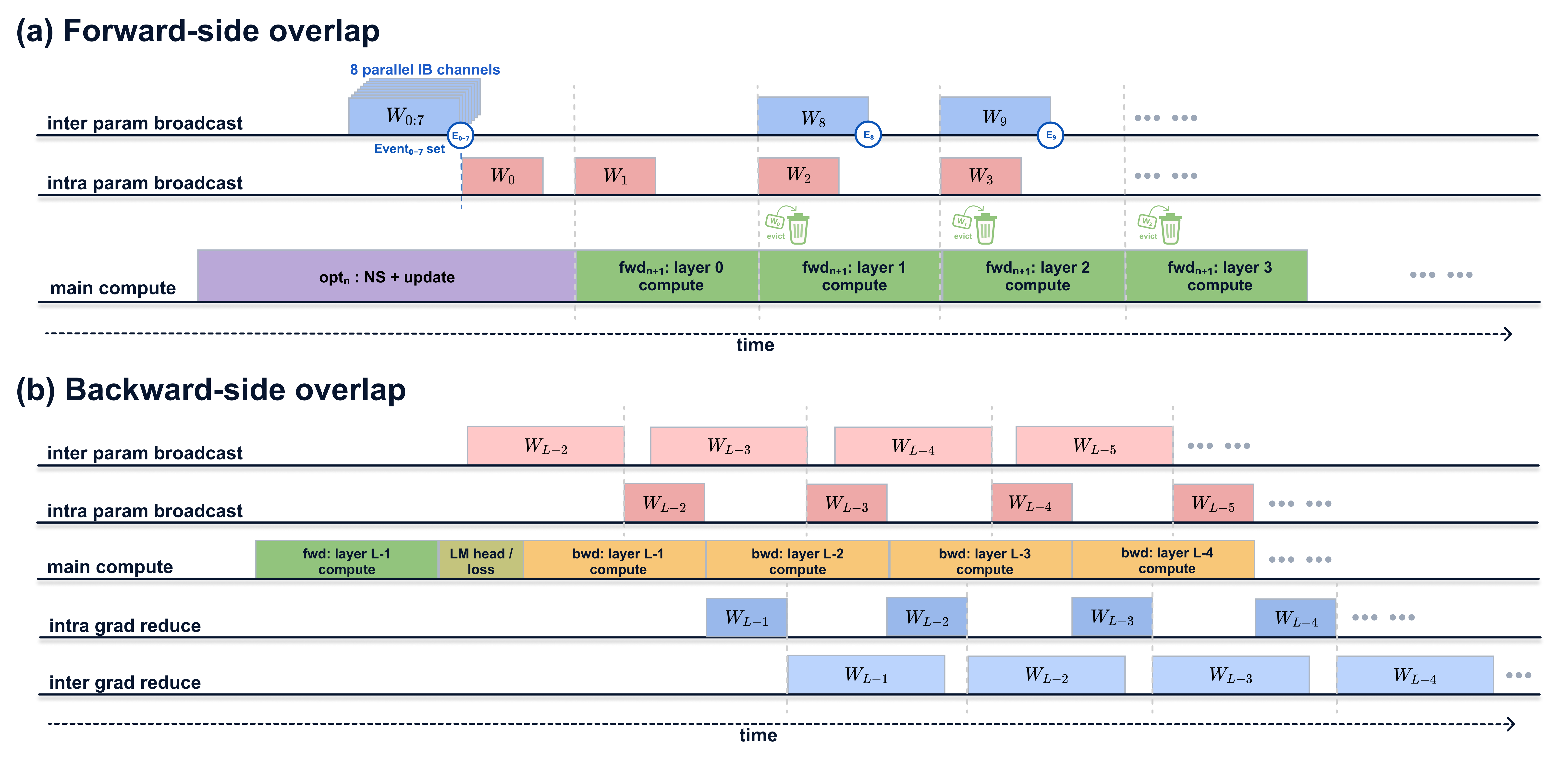}
\caption{\textbf{Forward and Backward pass}(a) In the forward pass, IB groups 0–8 perform inter-group broadcasts concurrently. Since these broadcasts are assigned to different groups, they do not interfere with each other. The gathering of the next weight shard is then triggered selectively, rather than by all groups simultaneously, to limit peak memory occupancy.
(b) In the backward pass, intra-group reduce operations may introduce communication interference. Therefore, we schedule the next broadcast before the reduce operation to avoid communication traffic contention. With the fine-grained weight layout, inter-group communication can be effectively overlapped.
}
\label{fig:fw_bp_overlap}
\end{figure}

\subsubsection{Forward Overlap}

\Cref{fig:fw_bp_overlap}(a)  shows the forward pass detail. Forward computation consumes parameters in layer order, but parameter
publication does not need to wait until the layer is reached. DMuon exploits
this gap by decoupling the two stages of publication. The inter-node stage,
which moves an owner-held weight to the corresponding GPU on each node, is
issued ahead of use over a lookahead window of layers. With the fine-grained
layout above, consecutive weights in this window are placed on different
inter-node columns, so their cross-node broadcasts can progress concurrently
with reduced collective-level contention.

The intra-node stage is delayed until the weight is close to consumption.
Once the inter-node copy of layer $l$ has arrived on each node, the runtime
broadcasts it locally across the node shortly before layer $l$ executes. This
creates a forward pipeline: while layer $l$ is computing, the runtime can
materialize layer $l+1$ within each node and simultaneously launch inter-node
publication for a later lookahead layer. The two communication stages use
different parts of the training mesh, allowing them to overlap with each other
and with forward compute.

Importantly, this overlap does not require abandoning the transient
materialization lifecycle that keeps peak memory low in sharded training.
Parameters are
materialized before the corresponding layer consumes them and released
once the layer finishes. As a result, the forward pipeline overlaps owner
publication with compute while retaining the peak-memory benefits of
FSDP-style parameter materialization, rather than requiring all owner-held
weights to be resident on every rank simultaneously.

\subsubsection{Backward Overlap}

\Cref{fig:fw_bp_overlap}(b) shows the backward pass detail. The backward execution contains two communication tasks around each layer. Before
a layer's backward computation can run, its parameters must be materialized
from their owners. After the computation finishes, the resulting weight
gradients must be routed back to the owners. These two tasks are ordered for
the same layer, but they are independent across adjacent layers in the
backward order. In particular, materializing the parameters for the next layer
to be executed does not depend on completing the gradient reduction of the
current layer.

DMuon exploits this independence by reordering backward communication into a
pipelined schedule. For each layer, the logical stages are

\[
\mathrm{bcast}_{\mathrm{inter}}
\;\rightarrow\;
\mathrm{bcast}_{\mathrm{intra}}
\;\rightarrow\;
\mathrm{compute}
\;\rightarrow\;
\mathrm{reduce}_{\mathrm{intra}}
\;\rightarrow\;
\mathrm{reduce}_{\mathrm{inter}} .
\]

The broadcast stages materialize the layer's parameters before backward
compute, while the reduce stages deliver the resulting gradients to the
owner. Across consecutive layers, these stages are overlapped: while the
gradient of layer $l$ is being reduced, the parameters of the next layer in
the backward order can already be broadcast. This preserves all data
dependencies, but avoids serializing parameter materialization and gradient
routing across the entire backward pass.

The schedule also respects communication contention. Intra-node broadcasts
and reductions share the same local fabric, so DMuon orders them to avoid
interference. Inter-node communication, however, is spread across the column
groups induced by the fine-grained weight layout, allowing broadcasts and
reductions from different layers to make progress concurrently. The resulting
pipeline keeps communication active on both sides of the backward computation
while delivering gradients to owners before the next Muon update.

\begin{figure}[b]
\centering
\includegraphics[width=\linewidth]{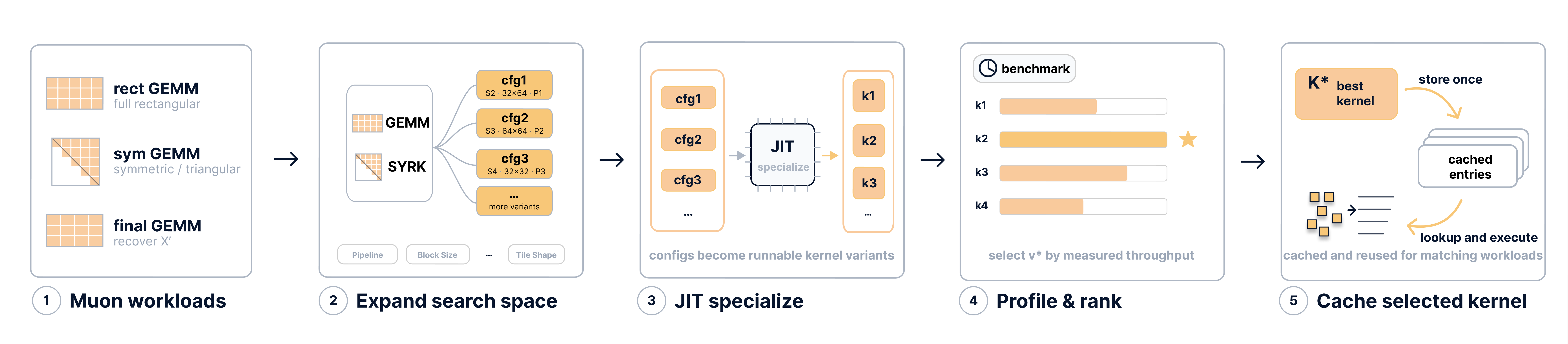}
\caption{
Autotuning workflow for DMuon's Gram NS kernels. Given a Muon workload, the runtime expands a search
space over tile shapes, block sizes, and software-pipeline configurations.
Tile-level DSLs generate specialized kernel variants, which are benchmarked
on the target hardware. The highest-throughput configuration is selected and
stored in a persistent kernel cache. Subsequent invocations with matching
workload characteristics bypass the search entirely and dispatch the cached
kernel directly.
}
\label{fig:syrk}
\end{figure}

\subsection{Efficient Gram Newton--Schulz Iteration}
\label{sec:design-gram-ns}

The owner-only strategy eliminates redundant Newton--Schulz (NS)
computation across ranks, but the owner-side orthogonalization remains the
dominant optimizer cost. To reduce it, DMuon adopts the Gram
Newton--Schulz formulation of \citet{daoGramNewtonSchulz2026}, which carries
the iteration entirely in Gram space.
Write the NS step as $X_{i+1}=P_iX_i$, with
$P_i := aI + bG_i + cG_i^2$ a polynomial in the Gram matrix
$G_i := X_iX_i^\top \in \mathbb{R}^{m\times m}$. Since $G_i$ is symmetric, so
is $P_i$, and the Gram matrix admits the closed recurrence

\begin{equation}
G_{i+1} = P_i\,G_i\,P_i .
\label{eq:gram-recur}
\end{equation}

The iteration thus stays in the $m\times m$ Gram space rather than the
original $m\times n$ space, reducing the dominant cost from $O(m^2n)$ to
$O(m^3)$ whenever $m<n$.

\begin{figure}[!ht]
\centering
\includegraphics[width=0.85\linewidth]{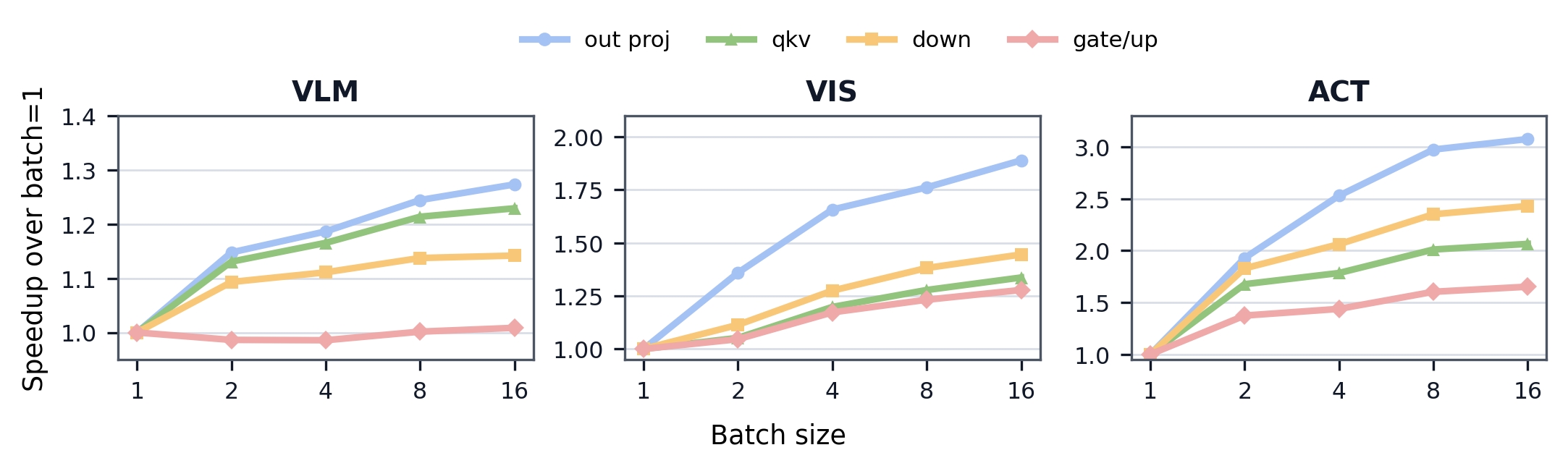}
\caption{Per-matrix time of the autotuned batched symmetric kernel, normalized
to single-matrix execution (batch~$=$~1), across the Gram-input shapes of
our workload. The amortization grows as matrices shrink: small,
near-square weights (right, e.g., $1024\!\times\!1024$) underfill the
device on their own and run up to $3\times$ faster per matrix at
batch~16, whereas large rectangular weights (left, e.g.,
$2048\!\times\!22016$) already saturate the GPU and gain little.}
\label{fig:batched-kernel}
\end{figure}

However, reducing arithmetic complexity alone does not guarantee efficient
execution in practice. Large matrices typically saturate the GPU and achieve high tensor-core
utilization, whereas many smaller matrices expose insufficient parallel work
to occupy all streaming multiprocessors.
This occupancy imbalance can be mitigated by a key property of the optimizer
workload, unlike the forward and backward passes, whose computation is
constrained by inter-layer dependencies, the Newton--Schulz iterations of
different weight matrices are entirely independent once gradients have been
reduced to their owners. DMuon exploits this independence through batched
execution (Fig.~\ref{fig:batched-kernel}): matrices that already saturate the device follow the standard
execution path, while matrices below an occupancy threshold are grouped by
shape and advanced through the Gram NS recurrence together as a single batched
iteration. This increases available parallelism, improves device utilization,
and amortizes kernel dispatch overhead across multiple matrices.

Beyond occupancy recovery through batching, the Gram-space formulation
exposes a second optimization opportunity through symmetry. The Gram matrix
$G=XX^\top$ is symmetric by construction, so computing all $m^2$ entries with
a general GEMM performs substantial redundant arithmetic. We therefore employ
a SYRK-style execution path that computes only the lower triangular portion of
$G$ in the mainloop and reconstructs the upper triangle in the epilogue. This
preserves the dense output required by subsequent Gram NS steps while nearly
halving the arithmetic work of the dominant Gram update. The symmetry-aware
implementation is shared by both the batched and non-batched execution paths.
Furthermore, the elementwise operations immediately following the Gram update
are fused into the same epilogue, allowing intermediate values to be consumed
before they are written back and reloaded from global memory. This eliminates
an additional kernel invocation and avoids a separate memory round-trip for
the elementwise stage.

Orthogonal to the occupancy and symmetry optimizations above, execution
efficiency remains highly shape dependent. Transformer training repeatedly
invokes only a small set of matrix shapes, but the performance-optimal kernel
schedule varies considerably across them. Differences in problem size and
aspect ratio can substantially change the best tile shape, software pipeline
depth, and warp-level schedule, making a single static configuration
suboptimal in practice.
To accommodate this shape diversity, DMuon treats kernel scheduling as a
shape-specific optimization problem(Fig.~\ref{fig:syrk}). We implement both the batched and
non-batched Gram NS execution paths using tile-level DSLs, including TileLang\cite{tilelang2025}
and CUTE DSL, which allow a large family of schedule variants to be generated
from a common kernel template. The autotuner explores a search space spanning
tile shapes, software pipeline depths, warp-level scheduling strategies, and
memory layouts, and evaluates candidate schedules directly on the target
hardware. When a matrix shape is encountered for the first time, DMuon
benchmarks the candidate schedules and selects the empirically fastest
configuration. The resulting schedule is recorded in a persistent kernel cache
keyed by problem shape and execution mode, subsequent invocations of the same
shape bypass the search entirely and dispatch the cached configuration
directly. This amortization strategy is particularly effective for optimizer
workloads, where the same parameter shapes recur throughout training, allowing
tuning cost to be paid once while retaining shape-specialized performance for
the later steps.

\subsection{Computation-Aware Load Balance}
\label{sec:loadbalance}

Under our parallel strategy, parameters are distributed across owner ranks at
matrix granularity. Because Muon update cost varies substantially with matrix
shape, the heterogeneous collection of weight matrices within a model leads
to uneven workloads across owners. Without careful assignment, some owners
become stragglers while others remain underutilized, causing the optimizer step time to be determined by the slowest rank.
A naive round-robin assignment ignores the substantial variation in execution
cost across matrix shapes. A more sophisticated approach is
Longest-Processing-Time (LPT), which balances workload according to an
analytical cost model. However, batching and autotuning introduce a much
richer execution space: the runtime of a Muon update depends not only on
matrix shape, but also on the selected batch size and kernel schedule. As a
result, execution cost can no longer be accurately characterized by a simple
analytical proxy.

Instead of relying on an approximate analytical model, owner assignment is
driven by a measured execution-cost model. Parameters are first grouped by
shape. Let $S$ denote the set of distinct matrix shapes and $n_s$ the number
of parameters of shape $s\in S$. Since model parameter shapes remain fixed
throughout training, the profiling process only needs to be performed once
during initialization. 
For each shape $s$, we evaluate a candidate set of batch sizes $B_s$ and
benchmark the owner-local Muon update under each configuration. The benchmark
includes the complete execution path selected by the runtime, including
batching behavior, kernel implementation, and autotuned schedules. Let
$c_{s,b}$ denote the measured execution time of processing one batch of shape
$s$ using batch size $b$. Unlike analytical estimates based on parameter count
or FLOPs, $c_{s,b}$ directly reflects the actual execution characteristics of
the target hardware.

Given the measured execution-cost model, owner assignment can be formulated as
a mixed-integer linear program (MILP) that minimizes the predicted makespan
across owner ranks. Let $R$ denote the set of owner ranks. For each shape
$s$, batch size $b\in B_s$, and owner $r\in R$, the integer variable
$x_{s,b,r}$ denotes the number of batches of that configuration assigned to
owner $r$. Using the measured execution costs $c_{s,b}$, the resulting
optimization problem is written as follows:

\begin{equation}
\begin{aligned}
\min \quad &
\max_{r\in R}
\sum_{s,b} c_{s,b}x_{s,b,r}
\\[1mm]
\mathrm{s.t.}\quad
&
\sum_{r,b}
b\,x_{s,b,r}
=
n_s,
\qquad
\forall s,
\\[1mm]
&
x_{s,b,r}\in\mathbb Z_{\ge0}.
\end{aligned}
\label{eq:owner_assignment}
\end{equation}

Minimizing the predicted makespan directly targets the optimizer-step critical
path. The equality constraint guarantees that all matrices of a given shape
are assigned exactly once.
The resulting MILP is solved once during initialization using SciPy's~\cite{virtanen2020scipy}
open-source MILP solver. Since model parameter shapes remain fixed throughout
training, the optimization is performed only once and the resulting
assignment is reused for all subsequent optimizer steps. In practice, the
one-time solve cost is negligible compared to the duration of training.

However, the MILP solve time grows with the size of the search space, which the number of distinct
shapes times the number of candidate owners. To bound the initialization
overhead, we impose a threshold $S_{\text{thr}}$ on the search-space size:
when the number of decision variables exceeds $S_{\text{thr}}$, we fall back
to a greedy search assignment instead of solving the MILP exactly. This preserves
the near-optimal balance of the MILP solver in the common regime while
guaranteeing a bounded, predictable initialization cost at large scale.

\subsection{The DMuon Training Step}
\label{sec:design-pipeline}

Algorithm~\ref{alg:dmuon-step} summarizes one logical DMuon training step on
a device mesh of size $G\!\times\!R$. A one-time setup
(lines~\ref{alg:line-lpt}--\ref{alg:line-alloc}) assigns matrix parameters to
owners and allocates owner-side parameter and optimizer state. Each training
step then proceeds through four phases: forward parameter materialization,
backward gradient routing, owner-side Muon update, and asynchronous
publication. Notation: $\ell$ indexes hook-boundary modules, typically
transformer blocks $(s^*_p,r^*_p)$ denotes the owner slot of parameter
$W^{(p)}$, $M^{(p)}$ is the owner-side Muon momentum, and
$\mathcal{S}_\text{bc}$ and $\mathcal{S}_\text{rd}$ denote the communication
streams described in \S\ref{sec:design-schedule}.

\begin{algorithm}[!ht]
\small
\caption{One DMuon training step on a device mesh of size $G\!\times\!R$.}
\label{alg:dmuon-step}
\begin{algorithmic}[1]
\Statex \textbf{Setup (once)}
\State $(s^*_p, r^*_p) \gets \mathrm{OwnerAssign}(\{W^{(p)}\})$
       \Comment{\S\ref{sec:loadbalance}}\label{alg:line-lpt}
\State Allocate $W^{(p)}$ and $M^{(p)}$ only on owner $(s^*_p, r^*_p)$
       \label{alg:line-alloc}

\Statex \textbf{Forward}
\For{$\ell = 1, \dots, L$}
  \State $\mathcal{S}_\text{bc}\!:$ wait $e_\text{pub}^{\ell}$;
         materialize $\{W^{(p)}\}_{p\in\ell} \to \mathrm{packed}_\ell$
         \label{alg:line-bc}
  \State forward($\ell$)
\EndFor

\Statex \textbf{Backward}
\For{$\ell = L, \dots, 1$}
  \State $\mathcal{S}_\text{bc}\!:$ materialize
         $\{W^{(p)}\}_{p\in\ell} \to \mathrm{packed}_\ell$
         \label{alg:line-bc2}
  \State backward($\ell$) $\to \mathrm{packed}_\ell.\mathrm{grad}$
  \State $\mathcal{S}_\text{rd}\!:$ reduce(\textsc{Avg},
         dst$=\!(s^*_p,r^*_p)$) over $G\!\times\!R$
         \label{alg:line-rd}
\EndFor

\Statex \textbf{Optimizer (owner-only)}
\ForAll{owned $W^{(p)}$}
  \State $O^{(p)} \gets \mathrm{GramNS}_k(M^{(p)})$
         \Comment{\S\ref{sec:design-gram-ns}}\label{alg:line-ns}
  \State $W^{(p)} \gets W^{(p)} - \eta O^{(p)}$
\EndFor
\State non-matrix params: sharded AdamW through the host stack

\Statex \textbf{Publish (async)}
\For{$\ell = 1, \dots, L$}
  \State $\mathcal{S}_\text{bc}\!:$ publish $\{W^{(p)}\}_{p\in\ell}$
         from owners; record $e_\text{pub}^{\ell}$
         \label{alg:line-pub}
\EndFor
\end{algorithmic}
\end{algorithm}

Algorithm~\ref{alg:dmuon-step} is written at the logical level, the
communication operations in lines~\ref{alg:line-bc}, \ref{alg:line-bc2},
\ref{alg:line-rd}, and~\ref{alg:line-pub} are implemented by the hierarchical
and pipelined schedule of \S\ref{sec:design-schedule}. The important semantic
property is that each owner receives the same averaged full-matrix gradient
$\bar{g}^{(p)}$ that a synchronous Muon reference would use, applies the same
momentum and Newton--Schulz update, and then publishes the updated parameter
for subsequent computation. Non-owner ranks materialize matrix parameters only
as temporary execution buffers and do not hold authoritative optimizer state
for them.

Forward and backward
parameter materialization are issued ahead of use and pipelined with
neighboring-layer computation. Gradient reductions are routed to owners while
the backward pass continues over earlier layers. Parameter publication is
decoupled across steps: after the owner update, updated weights are published
asynchronously, and the next step waits only at the pre-forward hook of the
layer that consumes the corresponding parameter. This organization preserves
exact Muon semantics while reducing the effective optimizer overhead seen on
the training critical path.       
\section{Implementation}
\label{sec:impl}

DMuon is implemented in approximately 10K lines of Python code with custom kernel sources, packaged as a standalone Python module whose
user-facing surface is three calls (\texttt{dedicate\_params},
\texttt{Muon}, and the state-dict accessors) and which preserves the
standard PyTorch optimizer protocol, and composes with a stock FSDP2 program with no source-level disruption.

\paragraph{Tensor-parallel composition.}
Tensor parallelism shards individual weight matrices across a TP group, so
a matrix assigned to a single DP owner slot is itself distributed over
several ranks. DMuon accommodates this by nesting a second level of
ownership within the first: after the cost-aware DP partition
(\S\ref{sec:loadbalance}) assigns each matrix to a DP owner slot,
a second pass over the matrices in that slot designates one rank of the TP
group as the \emph{TP owner} responsible for orthogonalizing it.
TP-sharded parameters are identified at \texttt{dedicate\_params} time by
inspecting each parameter's DTensor placements: any \texttt{Shard}
placement on a mesh dimension outside the DP mesh marks the parameter as
TP-sharded. We re-read this placement metadata at hook registration rather
than caching it, so that future changes to PyTorch's DTensor internals do
not break DMuon. The backward reduction delivers
gradients that remain TP-sharded, one slice per rank of the owner's TP
group, these slices are gathered at the TP owner, which assembles the full
gradient, runs the same Gram-space Newton--Schulz as in the non-TP case
(\S\ref{sec:design-gram-ns}), re-partitions the orthogonalized matrix, and
scatters each peer its update slice. From this point the step is
indistinguishable from the non-TP case: each rank applies its slice
locally, and the asynchronous publish broadcasts slices across the DP group
as before. TP handling is thus confined entirely to the optimizer step, the
forward broadcast, backward reduction, and asynchronous publish all operate
on the per-layer slices the host stack already maintains, and neither the TP
nor the DP code path is modified. Users compose TP, DMuon, and data-parallel
sharding in the same order they would with AdamW.

\paragraph{Non-owner placeholder.} On non-owner ranks, the original
parameters are replaced with a zero-size placeholder
carrying the same dtype. This preserves
all module-graph traversal code (Apex, PEFT, gradient clipping libraries
that walk \texttt{model.parameters()}) while consuming negligible
memory. The owner allocates a full-precision \texttt{\_owned\_data}
tensor on the same device. During unshard, a packed buffer is filled
from each owner's data and exposed back to user code
as a persistent \texttt{nn.Parameter} whose
storage is the packed buffer, autograd writes the gradient directly to
this parameter's \texttt{.grad} field, avoiding an intermediate copy.
Tied parameters (e.g., input embeddings aliased to an output head) are
replaced at every alias, so no alias escapes outside DMuon.

\paragraph{Ownership strategy plug-in.} The partition layer exposes
\texttt{load\_balance}, \texttt{round\_robin}, and \texttt{rank0} as named
strategies. The latter two exist for ablation: \texttt{rank0} forces a
single fixed owner (all matrices on rank~0) to show what happens if the
load-balance discipline is removed entirely. We include them as
ablation handles, not as production recommendations.

\paragraph{Configurations.} We adopt the Polar
Express~\cite{polarexpress2026} coefficient set as the default for
$k\!=\!5$ NS steps, with the standard $(a,b,c)$-quintic coefficients
selectable via configuration. The choice is orthogonal to DMuon's systems
contributions. We expose it because the symmetric kernel was tuned for the
matrix shapes Polar Express produces in its later steps.
A fine-grained analysis of the wgrad
distribution over the course of training shows that its dynamic range
stays well within what fp16 can represent. We therefore run the NS
iteration in fp16 rather than bf16, the two are identical in cost on
the tensor cores, but fp16's three additional mantissa bits give the
iteration higher precision at no latency penalty. Across the NS iteration everything stays in fp16 except the on-chip
accumulation. The orthogonalized update is cast to fp32 and applied to the fp32 master weights, and then cast back to the parameter's working dtype (bf16
in our deployments).         
\section{Evaluation}
\label{sec:eval}

Our evaluation is organized in two parts. We first present the main result in \ref{sec:eval-main}: an end-to-end, per-step time comparison of DMuon against AdamW and an unoptimized distributed Muon (Muon-AG) on four production workloads (Wall-OSS-05, Pi0, Wall-WM and Qwen2.5). We then report a speedup breakdown in \ref{sec:eval-breakdown} that attributes the observed speedup to each design component.

\paragraph{Setup.} All step-time measurements are collected on a A800-SXM4-80GB cluster (8 GPUs per node, NVLink within a node and 200 Gb/s InfiniBand across nodes) in bf16. The primary baseline is AdamW on the same setting. We additionally report Muon-AG, a vanilla distributed Muon that all-gathers the full gradient before computing the update.

\begin{table}[h]
\centering
\scriptsize
\setlength{\tabcolsep}{3pt}

\begin{tabular}{llrrrrrrrr}
\toprule
\multirow{2}{*}{Model} & \multirow{2}{*}{GPUs}
    & \multicolumn{2}{c}{DMuon}
    & \multicolumn{2}{c}{Vanilla}
    & \multicolumn{2}{c}{Speedup}
    & \multirow{2}{*}{\shortstack{AdamW\\Step}}
    & \multirow{2}{*}{$\Delta_{\text{A}}$} \\
\cmidrule(lr){3-4}
\cmidrule(lr){5-6}
\cmidrule(lr){7-8}
    & & Optim. & Step & Optim. & Step
    & Optim. & Step & & \\
\midrule
\multirow{6}{*}{Wall-OSS}
    & 8   & 112 & 1359 & 1693 & 2575 & 15.12 & 1.89 & 1324 & 2.7 \\
    & 16  & 71  & 1339 & 1758 & 2647 & 24.76 & 1.98 & 1328 & 0.7 \\
    & 32  & 43  & 1350 & 1754 & 2665 & 40.79 & 1.97 & 1342 & 0.9 \\
    & 64  & 29  & 1390 & 1798 & 2740 & 62.00 & 1.97 & 1364 & 1.9 \\
    & 128 & 19  & 1437 & 1851 & 2745 & 97.42 & 1.91 & 1412 & 1.8 \\
    & 256 & 18  & 1519 & 1977 & 2857 & 109.44 & 1.88 & 1496 & 1.5 \\
\midrule
\multirow{6}{*}{Pi0}
    & 8   & 80  & 1597 & 1124 & 2369 & 14.05 & 1.48 & 1498 & 6.6 \\
    & 16  & 46  & 1610 & 1123 & 2453 & 24.41 & 1.52 & 1545 & 4.1 \\
    & 32  & 30  & 1620 & 1147 & 2440 & 38.23 & 1.51 & 1581 & 2.5 \\
    & 64  & 25  & 1625 & 1179 & 2570 & 47.16 & 1.58 & 1595 & 1.9 \\
    & 128 & 16  & 1632 & 1240 & 2615 & 77.50 & 1.60 & 1616 & 0.6 \\
    & 256 & 14  & 1648 & 1308 & 2665 & 93.43 & 1.62 & 1637 & 1.2 \\
\midrule
\multirow{6}{*}{Wall-WM}
    & 8   & 472 & 2986 & 3707 & 6212 & 6.85  & 2.08 & 2539 & 17.6 \\
    & 16  & 246 & 2787 & 4047 & 6580 & 15.45 & 2.36 & 2590 & 7.6 \\
    & 32  & 181 & 2796 & 4046 & 6609 & 21.35 & 2.36 & 2615 & 6.9 \\
    & 64  & 135 & 2810 & 4248 & 6828 & 30.47 & 2.43 & 2685 & 4.7 \\
    & 128 & 99  & 2895 & 5105 & 7748 & 50.57 & 2.79 & 2794 & 3.6 \\
    & 256 & 64  & 3011 & 6265 & 9061 & 96.89 & 3.01 & 2915 & 3.3 \\
\midrule
\multirow{6}{*}{Qwen2.5-7B}
    & 8   & 214 & 2715 & 3469 & 5934 & 16.21  & 2.19 & 2590 & 4.8 \\
    & 16  & 121 & 2664 & 3528 & 6045 & 29.16  & 2.27 & 2602 & 2.4 \\
    & 32  & 79  & 2636 & 3511 & 6033 & 44.44  & 2.29 & 2624 & 0.5 \\
    & 64  & 46  & 2659 & 3526 & 6057 & 76.65  & 2.28 & 2643 & 0.6 \\
    & 128 & 32  & 2744 & 3554 & 6109 & 110.61 & 2.23 & 2734 & 0.4 \\
    & 256 & 22  & 2850 & 3604 & 6219 & 163.82 & 2.18 & 2844 & 0.2 \\
\bottomrule
\end{tabular}

\caption{Scaling validation across GPU counts (A800-80GB, bf16).
For each model we report the optimizer step time (\emph{Optim.}) and the
end-to-end per-step time (\emph{Step}) in ms, for DMuon and the vanilla
gather-then-compute distributed Muon. We additionally report the AdamW
end-to-end per-step time. Speedup is computed as vanilla divided by DMuon
for the optimizer time and the end-to-end step time, respectively.
$\Delta_{\text{A}}$ denotes the relative cost of DMuon compared with AdamW,
computed as $(\text{Step}_{\text{DMuon}}-\text{Step}_{\text{AdamW}}) /
\text{Step}_{\text{AdamW}} \times 100\%$.}
\label{tab:scaling}

\end{table}

\subsection{End-to-End Step Time}
\label{sec:eval-main}

Table~\ref{tab:scaling} compares the three optimizers on the four
workloads. The thesis is that DMuon brings Muon's per-step wall-clock
to within a few percent of AdamW. The gather-then-compute
\texttt{Muon-AG} baseline shows what a naive distributed Muon costs.

Across the evaluated workloads, DMuon remains close to AdamW in
end-to-end training throughput, with an average step-time overhead within
$+2\%$. Compared with the gather-then-compute \texttt{Muon-AG} baseline,
DMuon achieves a $1.48\times\!\text{--}\!3.01\times$ speedup in
end-to-end step time and a $6.85\times\!\text{--}\!163.00\times$
speedup in optimizer-step time. The speedup comes from replacing
global gather-then-compute execution with owner-side Newton--Schulz
updates that are distributed across ranks and overlapped with the
training step.

The remaining overhead relative to AdamW is not a scheduling artifact,
but the irreducible critical-path cost left after load balancing. It is
dominated by the Newton--Schulz time of the largest owner-side matrix
update, which must still be computed at least once for every Muon step.
Thus, DMuon removes the scalable part of Muon's distributed overhead, but
cannot eliminate the final largest NS update without changing the
optimizer itself. This makes DMuon a near-zero-overhead drop-in
replacement for AdamW while preserving the Muon update.

\begin{figure}[h!]
\centering

\begin{subfigure}[b]{0.49\linewidth}
    \centering
    \includegraphics[width=\linewidth]{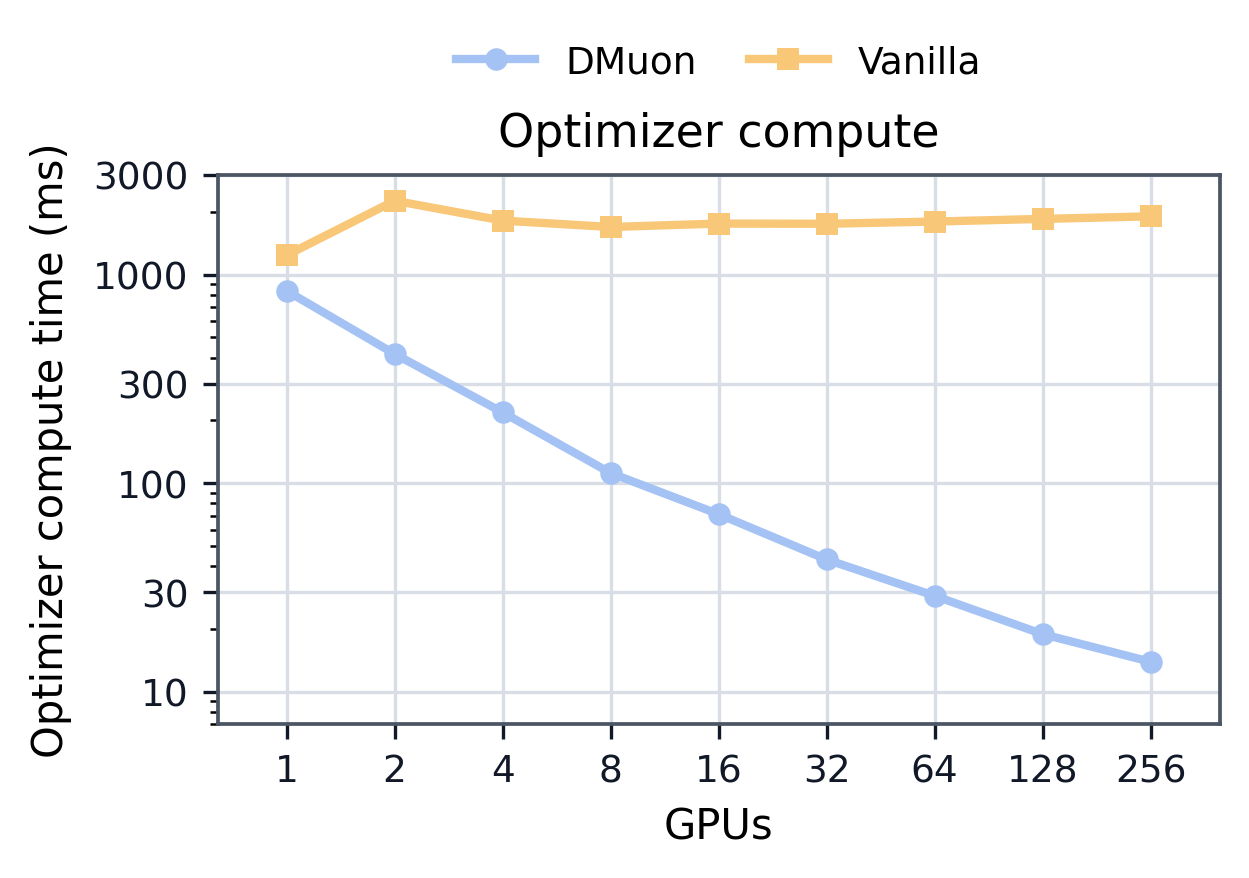}
    \caption{Optimizer compute time.}
    \label{fig:optimizer-compute-scaling}
\end{subfigure}
\hfill
\begin{subfigure}[b]{0.49\linewidth}
    \centering
    \includegraphics[width=\linewidth]{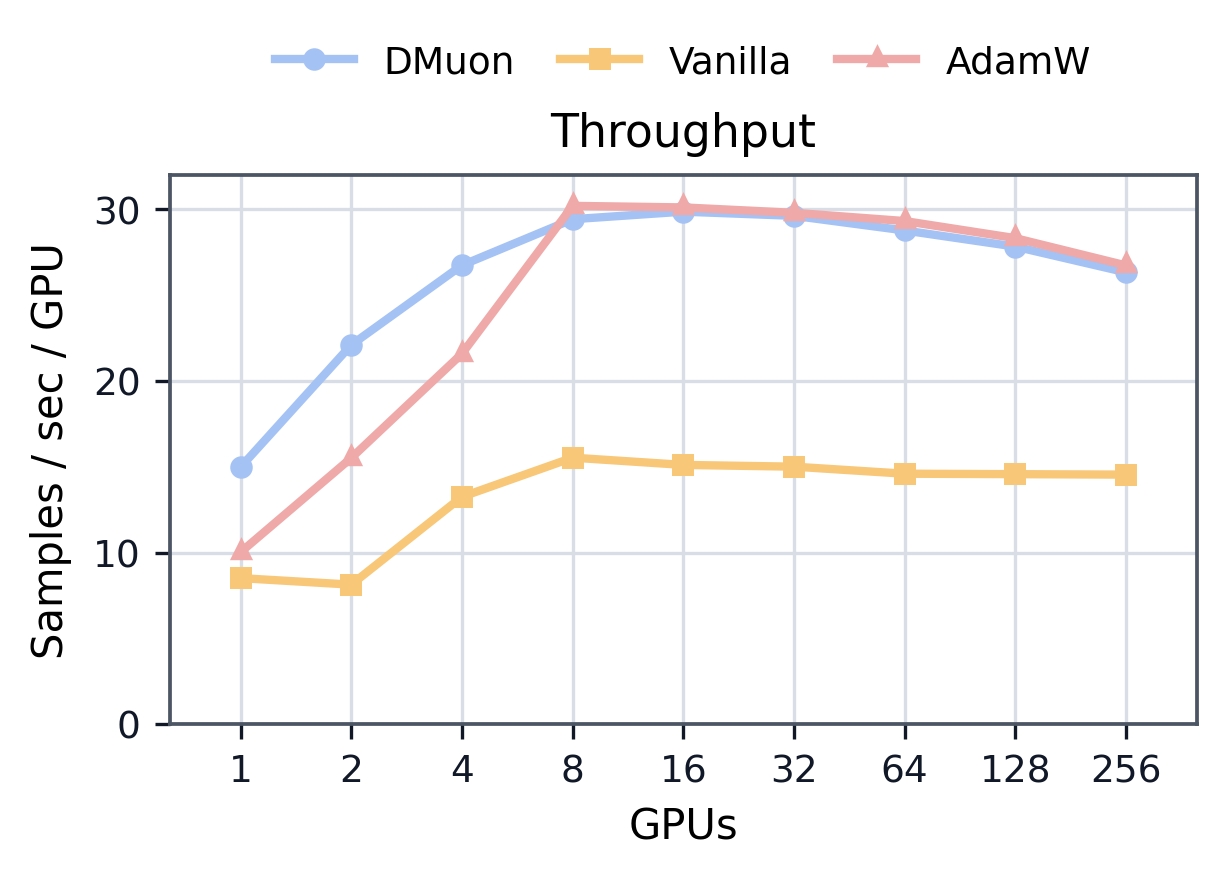}
    \caption{End-to-end training throughput.}
    \label{fig:throughput-scaling}
\end{subfigure}

\caption{
Scaling results on Wall-OSS across 1--256 A800 GPUs.
DMuon substantially reduces optimizer overhead and approaches AdamW-level throughput at larger scales.
}
\label{fig:scaling-results}
\end{figure}

Figure~\ref{fig:scaling-results} reports the scaling trend on Wall-OSS
from 1 to 256 A800 GPUs. At
smaller GPU counts, Muon-based training admits a larger batch size than AdamW
under the same memory budget, resulting in higher end-to-end throughput.
As distributed width increases and memory pressure is progressively reduced,
this batch-size advantage diminishes, causing DMuon throughput to gradually
converge toward the AdamW baseline.

\subsection{Speedup breakdown by component}
\label{sec:eval-breakdown}
Table~\ref{tab:ablation} attributes DMuon's optimizer-step speedup to
its three main components, measured by disabling each in isolation on
Wall-OSS-0.5 at 128~GPUs. The symmetric Gram kernel contributes the
largest share (48\%). Computing $G G^{\!\top}$ as a symmetric GEMM rather than
a general GEMM halves the arithmetic of the dominant products in
every NS iteration, and this saving applies to each of the five
iterations per matrix per step, making the kernel the single most
effective lever. Owner scheduling with load balancing assignment accounts for a
further 32\%: dedicating one owner per matrix removes the $D\times$
redundant NS computation, while load balancing prevents the largest
matrices from concentrating on a few ranks and turning into
stragglers that idle the rest of the group. The remaining 16\% comes
from auto-tuning and NS batching, which selects kernel configurations
per matrix shape and fuses the many small matrices of the model into
fewer launches, reducing per-launch overhead that would otherwise
dominate for sub-MFLOP problems. Together these components reduce the
end-to-end cost of running Muon to within 2\% of AdamW's step time on
average, effectively closing the gap between Muon's per-step price
and that of an element-wise optimizer.
\begin{table}[h!]
\centering
\small
\begin{tabular}{lr}
\toprule
Component & Share of speedup \\
\midrule
Symmetric Gram kernel        & 48\% \\
Owner scheduling \& load balancing  & 32\% \\
Auto-tuning \& NS batching          & 16\% \\
\midrule
\multicolumn{2}{l}{End-to-end step-time cost: \textbf{2\%} (avg. vs AdamW)} \\
\bottomrule
\end{tabular}
\caption{Per-component breakdown of DMuon's optimizer-step speedup. Each share is the fraction of the
total optimizer-time reduction attributable to that component.}
\label{tab:ablation}
\end{table}

\subsection{Limitations}
\label{sec:limitations}

DMuon is a set of mathematically equivalent reformulations of distributed Muon and does not change the update rule itself. It inherits Muon's convergence exactly and removes only systems overhead, not algorithmic cost.
A limitation of DMuon is that its benefit is less pronounced in
single-GPU training. While DMuon still reduces the optimizer-step time by
approximately $2\times$, it cannot leverage the additional benefits
provided by distributed execution.         
\section{Related Work}
\label{sec:relwork}
 
DMuon sits at the intersection of three lines of work: matrix-aware
optimizers, their distributed implementations, and hardware-aware
kernels for the Newton-Schulz (NS) iteration.
 
\paragraph{Matrix-aware optimizers.}
Shampoo~\cite{gupta2018shampoo} and SOAP~\cite{vyas2024soap}
precondition updates with Kronecker factors of the empirical Fisher.
Muon~\cite{jordan2024muon} instead applies an NS polar-factor to the
momentum-aggregated gradient, matching their quality with simpler
state. Recent variants NorMuon~\cite{normuon2025},
Newton--Muon~\cite{newtonmuon2026}, and Dion~\cite{dion2025} adjust
or low-rank-ify the orthogonalization. All share the same systems pain
point: their per-step update violates the element-wise contract that
ZeRO and FSDP assume. DMuon addresses this systems problem. it does
not propose a new optimizer.
 
\paragraph{Muon at scale.}
Moonlight~\cite{moonlight2025} established that Muon scales (a 16B MoE
trained on 5.7T tokens at $\sim\!2\times$ AdamW's token efficiency),
and Kimi-K2~\cite{kimi-k2} confirmed the result at trillion-parameter
scale. Gather-then-compute implementations
(\S\ref{sec:bg-mismatch}) in which every rank all-reduces the full
gradient and runs NS redundantly, a strategy our measurements
(Table~\ref{tab:scaling}) show carries a
significant step-time penalty over AdamW.
 
\paragraph{Distributed matrix optimizers.}
Distributed Shampoo~\cite{shi2023distshampoo} introduced an
owner-compute/all-gather paradigm for distributed matrix optimization under data-parallel training: the memory and computation of Shampoo updates are assigned to workers, and the resulting search directions are all-gathered at each step. Canzona~\cite{canzona2026} addresses a closely related problem in Megatron-style training stacks using an
$\alpha$-balanced data-parallel partition and tensor-parallel micro-group scheduling, reporting a $1.57\times$ end-to-end speedup on Qwen3-32B at 256 GPUs. In contrast, DMuon realizes owner-only Muon as a fine-grained overlap runtime for PyTorch FSDP2/HSDP's ZeRO-3-style execution, using scheduled broadcasts, two-stage reductions, and asynchronous update publishing to hide communication costs and alleviate peak memory pressure.

\paragraph{Sharded training stacks.}
ZeRO~\cite{rajbhandari2020zero} partitions optimizer state, gradients,
and parameters across DP ranks. FSDP~\cite{zhao2023fsdp} is its
PyTorch-native realization, with FSDP2's DTensor representation
adopted in modern stacks~\cite{torchtitan2024} and HSDP arranging
ranks on a 2-D mesh to bound the all-gather radius. DMuon operates against these sharded-training contracts while preserving their external execution model, requiring no source-level modifications to the host training framework.

\paragraph{Hardware-aware Newton-Schulz.}
Gram Newton-Schulz~\cite{daoGramNewtonSchulz2026} recasts each NS step
as a recurrence in the Gram matrix $XX^\top$, reducing the work to one
symmetric product plus a polynomial in the smaller Gram matrix. Polar
Express~\cite{polarexpress2026} optimizes per-step coefficients for
fixed iteration counts. DMuon adopts both as defaults and contributes
a shape-specialized symmetric kernel tuned for the matrix shapes of production LLM and VLA transformers.

\paragraph{Automatic kernel generation.}
A complementary line of work synthesizes high-performance kernels
automatically. Tensor compilers such as TVM~\cite{chen2018tvm} search
over loop-level schedules, with AutoTVM~\cite{chen2018autotvm} learning
a cost model over hand-written templates,
Ansor~\cite{zheng2020ansor} and FlexTensor~\cite{zheng2020flextensor}
generating the search space itself, and Tensor
Comprehensions~\cite{vasilache2018tensorcomprehensions} deriving
schedules from a polyhedral model. Closer to the hardware, tile-level
programming languages Triton~\cite{tillet2019triton} and
TileLang~\cite{tilelang2025}, expose tiles as first-class objects so
that near-peak kernels can be written without descending to raw
CUDA/PTX. DMuon's kernel implementation (\S\ref{sec:design-gram-ns}) is inspired by these systems. 

\section{Conclusion}
\label{sec:conclusion}

We presented DMuon, a distributed runtime and kernel stack for scaling
the Muon optimizer on modern sharded-training systems. DMuon combines
fine-grained communication optimization, computation-aware load balance, and high-performance kernel system to eliminate redundant optimizer
computation and overlap the remaining work with training execution. DMuon maintains end-to-end step times within
$2\%$ of AdamW on average, making Muon a practical drop-in optimizer for
large-scale production training.

\section*{Contributors}
\label{sec:contributors}

DMuon is a collaborative effort of the X Square Robot team. The full contributor list is given below; $^{\ast}$ denotes core contributors, $^{\dagger}$ denotes the project lead, and $^{\ddagger}$ denotes the corresponding author.

\vspace{0.5em}
\noindent
Vincent Chen$^{\ast\dagger}$, Starrick Liu$^{\ast}$, Regis Cheng$^{\ast}$, Dance Yang$^{\ast}$, Shalfun Li$^{\ast}$, Ryan Yu, Lucy Liang, Hang Su, Roy Gan, Hao Wang$^{\ddagger}$, Qian Wang.

\clearpage
\newpage
\bibliographystyle{assets/plainnat}
\bibliography{paper}

\end{document}